\documentclass[%
notitlepage,
superscriptaddress,
amsmath,amssymb,
aps,
pra,
twocolumn,
floatfix,
]{revtex4-2}

\usepackage{tcolorbox}
\usepackage{graphicx}% Include figure files
\usepackage{dcolumn}% Align table columns on decimal point
\usepackage{bm}% bold math
\usepackage{nicefrac}       % compact symbols for 1/2, etc.
\usepackage{amsthm}
\usepackage{amsmath}
\usepackage{amsfonts}
\usepackage{amssymb}
\usepackage[columnwise]{lineno}
\usepackage{siunitx}

\makeatother
\usepackage{hyperref}% add hypertext capabilities
\hypersetup{
    colorlinks=true,
    linkcolor=blue,
    filecolor=magenta,      
    urlcolor=cyan,
}% add hypertext capabilities

\usepackage{color}

\usepackage[normalem]{ulem}

\begin{document}
%%%%%%%%%%%%%%%%%%%%%%%%%%%%%%%%%%%%%%%%%%%%%%%%%%%%%%%%%%%%%
\title{Ray-based framework for state identification in quantum dot devices}

\author{Justyna P. Zwolak}
\email{jpzwolak@nist.gov}
\affiliation{National Institute of Standards and Technology, Gaithersburg, MD 20899, USA}

\author{Thomas McJunkin}
\email{tmcjunkin@wisc.edu}
\affiliation{Department of Physics, University of Wisconsin-Madison, Madison, WI 53706, USA}

\author{Sandesh S. Kalantre}
\affiliation{Joint Quantum Institute, University of Maryland, College Park, MD 20742, USA}
\affiliation{Joint Center for Quantum Information and Computer Science, 
University of Maryland, College Park, MD 20742, USA}

\author{Samuel F. Neyens}
\affiliation{Department of Physics, University of Wisconsin-Madison, Madison, WI 53706, USA}

\author{E.~R.~MacQuarrie}
\affiliation{Department of Physics, University of Wisconsin-Madison, Madison, WI 53706, USA}

\author{Mark A. Eriksson}
\affiliation{Department of Physics, University of Wisconsin-Madison, Madison, WI 53706, USA}

\author{Jacob M. Taylor}
\affiliation{National Institute of Standards and Technology, Gaithersburg, MD 20899, USA}
\affiliation{Joint Quantum Institute, University of Maryland, College Park, MD, 20742 USA}
\affiliation{Joint Center for Quantum Information and Computer Science, University of Maryland, College Park, MD 20742, USA}

\date{\today}
%%%%%%%%%%%%%%%%%%%%%%%%%%%%%%%%%%%%%%%%%%%%%%%%%%%%%%%%%%%%%
%%%%%%%%%%%%%%%%%%%%%%%%%%%%%%%%%%%%%%%%%%%%%%%%%%%%%%%%%%%%%
\begin{abstract}
Quantum dots (QDs) defined with electrostatic gates are a leading platform for a scalable quantum computing implementation. 
However, with increasing numbers of qubits, the complexity of the control parameter space also grows. 
Traditional measurement techniques, relying on complete or near-complete exploration via two-parameter scans (images) of the device response, quickly become impractical with increasing numbers of gates. 
Here we propose to circumvent this challenge by introducing a measurement technique relying on one-dimensional projections of the device response in the multidimensional parameter space. 
Dubbed the ``ray-based classification (RBC) framework,'' we use this machine learning approach to implement a classifier for QD states, enabling automated recognition of qubit-relevant parameter regimes. 
We show that RBC surpasses the $82\,\%$ accuracy benchmark from the experimental implementation of image-based classification techniques from prior work, while reducing the number of measurement points needed by up to $70~\%$. 
The reduction in measurement cost is a significant gain for time-intensive QD measurements and is a step forward toward the scalability of these devices. 
We also discuss how the RBC-based optimizer, which tunes the device to a multiqubit regime, performs when tuning in the two-dimensional and three-dimensional parameter spaces defined by plunger and barrier gates that control the QDs.
This work provides experimental validation of both efficient state identification and optimization with machine learning techniques for non-traditional measurements in quantum systems with high-dimensional parameter spaces and time-intensive measurements.
\end{abstract}

\maketitle
%%%%%%%%%%%%%%%%%%%%%%%%%%%%%%%%%%%%%%%%%%%%%%%%%%%%%%%%%%%%%
%\linenumbers
%%%%%%%%%%%%%%%%%%%%%%%%%%%%%%%%%%%%%%%%%%%%%%%%%%%%%%%%%%%%%
\section{Introduction}
%%%%%%%%%%%%%%%%%%%%%%%%%%%%%%%%%%%%%%%%%%%%%%%%%%%%%%%%%%%%%
The ease of control~\cite{Petta05-CQM, Koppens06-COS, Medford13-QEQ,Kim15-RGQ}, fast measurement~\cite{Barthel09-SSM}, and long coherence~\cite{Veldhorst14-AQD, Kawakami14-LLQ, Yoneda18-QDC} of semiconductor quantum dots (QDs) make them a promising platform for quantum computing~\cite{Zwanenburg13-SQE}. 
Individual qubits can be built from single QDs~\cite{Loss98-QCD} or multiple QDs coupled together~\cite{Petta05-CQM, Shi2012QDHQ, Laird2010XO}. 
At present, most QD qubit systems require multiple electrostatic gates to isolate, control, and sense each qubit. 
Often, there are specific gates designed to accumulate electrons into QDs (plungers), gates to control the tunneling between QDs (barriers), and gates to deplete electrons elsewhere (screening gates)~\cite{Angus07-GDS}. 
As QD devices grow in the number of qubits~\cite{Zajac16-SGA} and complexity~\cite{Mukhopadhyay18-2DD}, so do the number of gate voltages to be controlled and tuned. 

Although current few-qubit devices~\cite{Watson18-PQP, Zajac18-RDC, Hendrickx2020} are mostly still tuned manually, there are several emerging automated approaches to various steps in the process of tuning QDs. 
Depending on the specific device design, each of these tuning steps requires specialized approaches for automation. 
Some automation techniques focus on tuning devices \textit{ab initio} to a voltage space where QDs can form~\cite{Darulova2020, Moon2020}. 
Others focus on tuning the configuration of QDs; that is from single QDs to coupled double QDs~\cite{Zwolak20-AQD}. 
There are also methods to achieve a specific number of electrons in each QD~\cite{Durrer2020} or to measure and modify the couplings in multiple-QD systems~\cite{Mills19-CAT, van_Esbroeck2020}. 
These various automation techniques have used many different tools: convolutional neural networks (CNNs)~\cite{Kalantre17-MLD, Zwolak20-AQD}, deep generative modeling~\cite{Lennon19-EMM}, classical feature extraction (e.g., a Hough transformation)~\cite{Mills19-CAT, Lapointe2020}, and many custom fitting models~\cite{Baart2016-ATD}. 

\begin{figure*}[t]
\includegraphics[width=1.0\linewidth]{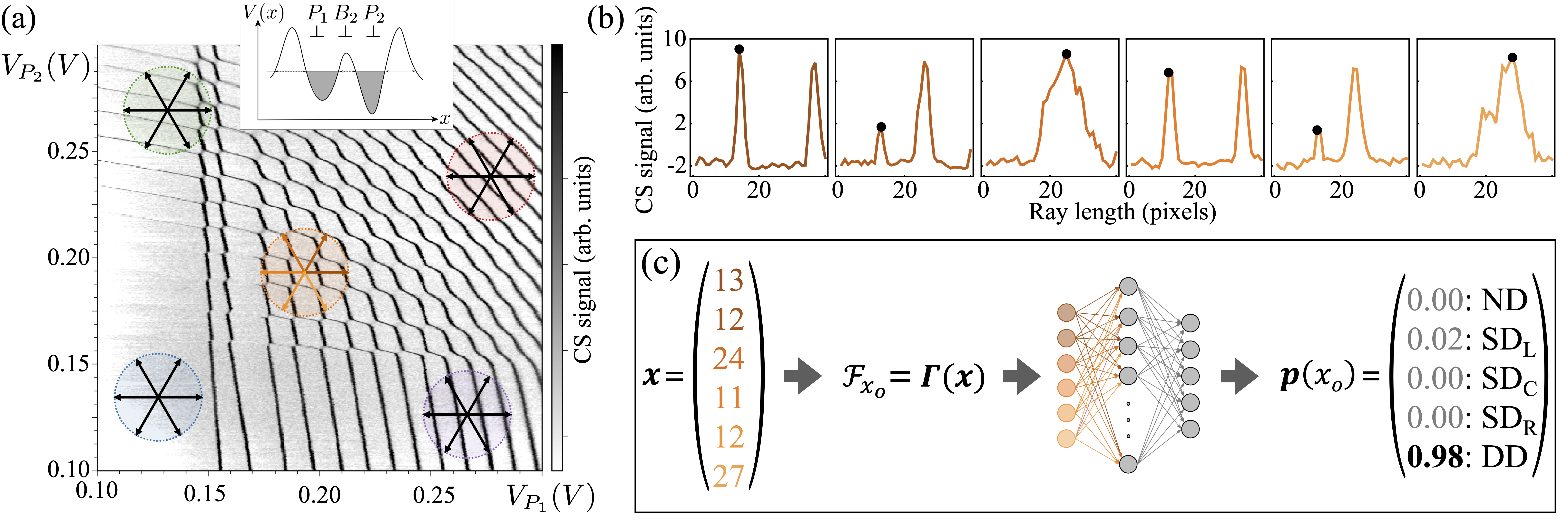}
\caption{Visualization of the ray-based fingerprinting framework. 
(a) A preview of five points in the space of plunger gates $(P_1,P_2)$ with six evenly distributed rays overlaying a sample measurement. 
Different colors represent different QD states. The inset shows the potential landscape of the double-quantum-dot system. 
Gate voltages $V_{P_1}$ and $V_{P_2}$ control the occupation of each QD.
Gate voltage $V_{B_2}$ controls the inter-QD tunneling.
(b) The preprocessed charge sensor (CS) signal for six evenly distributed rays measured from a point $(V_{P_1},V_{P_2})=(0.193,0.193)\,$\si{\volt} [the most central point in (a)] as a function of ray length. 
The length resolution is $0.5\,$\si{\milli\volt} per pixel.
In each ray, the position of the transition line nearest to the point $(V_{P_1},V_{P_2})\equiv x_o$---that is, the critical feature along a given ray---is marked with a dot. 
(c) The flow of the RBC framework adapted from Ref.~\cite{Zwolak20-RBC}. A vector of critical features $\bm{x}$ is processed using a weight function $\Gamma(\bm{x})$. 
The resulting fingerprint $\mathcal{F}_{x_o}$ is processed by a DNN classifier, returning a probability vector $\bm{p}(x_o)$ quantifying the current state of the device at point $x_o$.}
\label{fig:rbc_overview}
\end{figure*}

Motivated by the success of image-based autotuning~\cite{Zwolak20-AQD}, here we present an alternative approach that uses the recently proposed ray-based classification (RBC) framework~\cite{Zwolak20-RBC} to distinguish between different electron configurations.
The RBC framework was originally proposed as an approach for classifying simple bounded and unbounded convex geometrical shapes. 
It thus naturally applies to identifying QD states that manifest themselves as distinct geometrical patterns in the charge sensor response as a function of the gate voltages.
Here we present the classification of a Si/Si$_x$Ge$_{1-x}$ QD device~\cite{Neyens2019} using this new method, both in a ``live'' measurement session during the experiment and ``off-line'' using a dataset of large stability diagrams taken from the device after tuning.

We explore how the hyperparameters of the RBC, such as number of rays, ray length, and the choice of the weight function, affect the classification accuracy of experimental data. We find a favorable comparison with image-based classification in terms of accuracy and the quantity of data required. 
Furthermore, we show an off-line implementation of the RBC framework within an optimizer-based autotuner for a QD system, tuning between single and double QDs in a space of three gate voltages.

The paper is organized as follows: In Sec.~\ref{sec:framework}, we describe the framework for ray-based classification in the context of QD devices as well as the experimental setup and characteristics of the QD chip used in the experiment. 
The performance of the classifier and the off-line implementation of the autotuning protocol are discussed in Sec.~\ref{sec:results}. 
We conclude with a discussion of the potential modifications to further improve the proposed autotuning technique in Sec.~\ref{sec:conclusion}.

%%%%%%%%%%%%%%%%%%%%%%%%%%%%%%%%%%%%%%%%%%%%%%%%%%%%%%%%%%%%%
\section{Ray-based classification framework}\label{sec:framework}
%%%%%%%%%%%%%%%%%%%%%%%%%%%%%%%%%%%%%%%%%%%%%%%%%%%%%%%%%%%%%
A visual inspection of the large scan of experimental data (differential charge sensing) presented in Fig.~\ref{fig:rbc_overview}(a) shows different physical states of the QD device. 
These states manifest themselves as different shapes formed by electron transition lines and varying orientations with respect to the scanned gate voltages (e.g., parallel lines for single QDs and honeycombs for double QDs). 
Thus, the shape and orientation of the lines encode sufficient qualitative information about the state of the device to enable state (in this case, charge topology) classification. 
As shown in Ref.~\cite{Ziegler21-RAN}, a CNN-based classifier trained for state identification learns to mask the noise captured between transition lines in these two-dimensional (2D) charge sensing images. 

Recently, a classification framework focusing on data acquisition efficiency was proposed by Zwolak {\it el al.}~\cite{Zwolak20-RBC}. 
Rather than using full 2D images capturing a small region of the voltage space, the RBC framework relies on a collection of evenly distributed one-dimensional traces (``rays'') originating from a single point $x_o$ and measured in multiple directions in the voltage space to describe the neighborhood of $x_o$ (see Fig.~\ref{fig:rbc_overview}(a) for a preview of five sample points with six evenly distributed rays).
The rays are used to capture the orientation and relative position of transition lines near $x_o$, effectively ``fingerprinting'' the surrounding voltage space.
The resulting point fingerprint encodes the qualitative information about the voltage space around $x_o$ and is the primary object of the RBC framework. 
The full mathematical formalism of the framework is presented in Ref.~\cite{Zwolak20-RBC}.

%%%%%%%%%%%%%%%%%%%%%%%%%%%%%%%%%%%%%%%%%%%%%%%%%%%%%%%%%%%%%
\subsection{Experimental realization}\label{sec:experiment}
%%%%%%%%%%%%%%%%%%%%%%%%%%%%%%%%%%%%%%%%%%%%%
A Si/Si$_x$Ge$_{1-x}$ quadruple-QD device as described in Ref.~\cite{Neyens2019} is used to create a double-QD charge sensed by a single sensing QD whose current readout is connected to a cryogenic amplifier. 
The device is a linear array of four QDs, opposing two charge sensors (see Fig. 1(a) in Ref.~\cite{Neyens2019}). 
The nearby gates (reservoir gates, depletion gates, and tunnel-barrier gates) are pretuned to allow single-QD and double-QD formation under the two leftmost plunger gates, $P_1$ and $P_2$ (see the inset in Fig.~\ref{fig:rbc_overview}(a)). 
Further device details can be found in Ref.~\cite{Neyens2019}.
An example stability diagram for this device is shown in Fig.~\ref{fig:rbc_overview}(a). 
A small, approximately-10-\si{\kilo\hertz} oscillating voltage is applied to $P_1$ and the charge sensor current is sent to a lock-in amplifier referenced to this ac tone. 
This results in a large change in the signal measured at charge transitions, an effective differentiation of the QD occupation across the $(P_1, P_2)$ voltage space. 
Because the ac tone is applied to $P_1$, charge transitions physically closer to $P_1$ will result in a larger signal than transitions closer to $P_2$. 
This effect can be seen in Fig.~\ref{fig:rbc_overview}(a), where the more horizontal transitions associated with occupation changes in the $P_2$ QD are harder to distinguish. 
In future measurements, this effect could be reduced by also applying an ac tone to $P_2$ or applying the tone to a central tunnel barrier gate.

To assess the geometry of the transition lines surrounding a given point $x_o\in(V_{P_1},V_{P_2})$, we consider a collection of $M$ rays of a fixed length centered at $x_o$ called the $M$-projection (see Fig.~\ref{fig:rbc_overview}(a) for visualization). 
Each ray corresponds to a measurement of the charge sensor signal along a given direction in the space of plunger voltages.
The ray data used in this paper are collected in two ways. 
The ``live'' $M$-projection is collected by choosing a plunger gate voltage point $x_o=(V_{P_1},V_{P_2})$ and measuring evenly spaced rays emanating from that point in the plunger gate voltage space.
The length of the rays and their granularity (i.e., number of pixels per unit length) are determined by the expected charging energy of the system and are fixed throughout the measurement. 
We use rays 30~\si{\milli\volt} in length with 60 points (pixels) sampled along the ray. 
The 0.5~\si{\milli\volt}-per-pixel granularity is selected to ensure that the electron transition lines will be be properly visible with the ac lock-in measurement technique.
For the ``off-line'' $M$-projection, a large, densely sampled 2D stability diagram is used to generate ray datasets by choosing a central voltage point $x_o$ and interpolating the data in evenly spaced directions. 
In both cases, the first ray is always measured in the direction of $V_{P_1}$.

Regardless of the ray data generation method, we collect complex voltage data from the lock-in amplifier.
Fig.~\ref{fig:rbc_overview}(b) shows the magnitude of a set of live data rays. 
In the off-line setting, a combination of the overall median absolute deviation and the median for a given collection of rays is used to determine the noise level and expected peak prominence, respectively, and is used by the peak finding algorithm.
In an {\it in situ} implementation, the noise level can be determined before ray collection by measuring the average lock-in response offset and rms noise at any off-transition plunger voltage and then periodically rechecked throughout the experiment.

Once an $M$-projection for a given point $x_o$ is acquired, traditional signal processing techniques are used to test each ray for the presence of transition lines.
While the noiseless simulation results in binary rays, with transitions easily identifiable along the rays, the noise present in the experimental data makes the transitions harder to detect.
In the ac measurement, transitions manifest themselves as peaks along the ray (called ``features'' in the RBC framework; see Fig.~\ref{fig:rbc_overview}(b)).
Thus, a peak detection algorithm is applied to each ray to determine the presence and, if applicable, positions of all peaks along a given ray. 
If a dc charge sensor measurement is used instead, an additional step of differentiating the signal along the measurement direction will be necessary before signal processing. 
The peak positions are represented as a number of pixels from the central voltage point $x_o$.
If for a given ray at least one feature is detected, the position of the feature nearest to the ray's origin $x^{(c)}$ is recorded (so-called critical feature).
If no peaks are found, a not-a-number (NaN) value is recorded as a placeholder for the critical feature instead.
The vector of critical features $\bm{x}$, marked with black points in Fig.~\ref{fig:rbc_overview}(b), is used to determine the point fingerprint.

Finally, a ``weight'' function $\bm{\Gamma}$ is applied elementwise to scale the vector of critical features to a $[0,1]$ range, with rays having no peaks being assigned a default value of $0$:
\begin{equation}
    \bm{\Gamma}(\bm{x}) = 
    \begin{cases} \gamma(x_i^{(c)}) &\text{if } x_i^{(c)} \in \mathbb{N}^{>0},\\
    0 &\text{if } x_i^{(c)} = \rm{NaN}, \end{cases}
\end{equation}
where $\gamma:\mathbb{N}^{>0}\rightarrow[0,1]$ is a normalizing decreasing function.
The normalized vector of distances $\mathcal{F}_{x_o}$ is called the ``point fingerprint''. 
Because of the differences in the geometry of the transition lines for different QD states, distinct point fingerprints are encoded for the different states and a classifier trained on point fingerprint data suffices for the QD state identification. 
We use a simple deep neural network (DNN) classifier with three hidden layers for this purpose.  

The flow of the RBC algorithm is shown in Fig.~\ref{fig:rbc_overview}(c). It consists of three main steps: 
\begin{enumerate}
    \item Extraction of the positions of critical features from the $M$-projection,
    \item Fingerprinting of the central point $x_o$ by the means of a weight function $\gamma(x)$.
    \item DNN analysis of the resulting fingerprint $\mathcal{F}_{x_o}$.
\end{enumerate}
The output of the classifier is a probability vector,
\begin{equation}\label{eq:prob_vec}
\bm{p}(x_o)=[p_{\rm ND},\,p_{{\rm SD}_L},\,p_{{\rm SD}_C},\,p_{{\rm SD}_R},\,p_{\rm DD}]
\end{equation} 
quantifying the current state of the device, with ND denoting no QDs formed, ${\rm SD}_L$, ${\rm SD}_C$, and ${\rm SD}_R$ denoting the left, central, and right single QD, respectively, and ${\rm DD}$ denoting the double-QD state.

\begin{figure*}[t]
\includegraphics[width=1.0\linewidth]{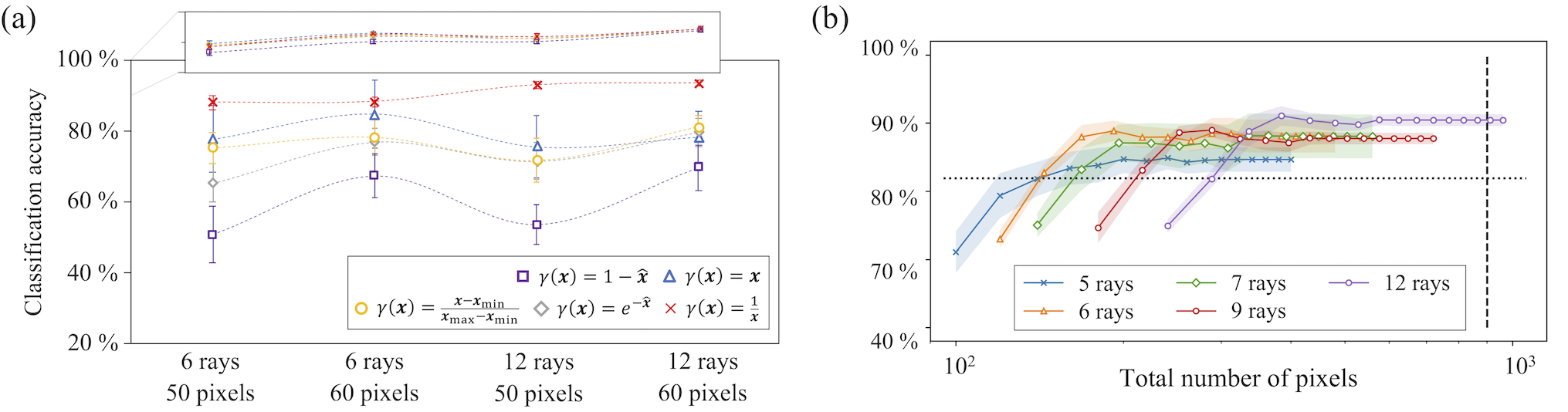}
\caption{(a) Classifier performance for various weight functions $\gamma(x)$ and for different numbers of rays and ray lengths using off-line experimental data (averaged over $N=20$ models). 
The inset at the top shows the performance on simulated data. 
(b) Classifier performance for different numbers of rays as a function of the total number of pixels measured for off-line experimental data (averaged over $N=20$ models). 
The dotted black horizontal line represents the benchmark from Ref.~\cite{Zwolak20-AQD}, while the dashed vertical line represents the minimum data requirement for a CNN classifier as used in Ref.~\cite{Zwolak20-AQD}. 
In both panels, the connecting lines are a guide for the eye only. The error bars (a) and bands (b) correspond to one standard deviation.}
\label{fig:perf_comparison}
\end{figure*}

%%%%%%%%%%%%%%%%%%%%%%%%%%%%%%%%%%%%%%%%%%%%%%%%%%%%%%%%%%%%%
\section{Results}\label{sec:results}
%%%%%%%%%%%%%%%%%%%%%%%%%%%%%%%%%%%%%%%%%%%%%%%%%%%%%%%%%%%%%
The RBC framework was developed and tested originally on a dataset of simulated double-QD devices~\cite{Zwolak20-RBC}.
In that work, an average accuracy of $96.4(4)~\%$ (averaged over $N=50$ models) with just six rays and a weight function $\gamma(x)=1/x$ was reported for double QDs, where the accuracy is defined as the fraction of correctly classified points from a test dataset.
This is on par with the more-data-demanding CNN-based classification framework, while requiring $60~\%$ fewer data~\cite{Zwolak18-QLD, Zwolak20-RBC}.
Given the success of the RBC framework on simulated devices, here we focus on its performance on experimental data, where the reduction in the data required directly translates to reduction of the measurement time in the experiment.

%%%%%%%%%%%%%%%%%%%%%%%%%%%%%%%%%%%%%%%%%%%%%%%%%%%%%%%%%%%%%
\subsection{Ray based classification}\label{subsec:rbc}
%%%%%%%%%%%%%%%%%%%%%%%%%%%%%%%%%%%%%%%%%%%%%
To assess the performance of the RBC framework with experimental data, we use an ensemble of 20 DNN classifiers pretrained using a modified version of the ``Quantum dot data for machine learning'' dataset~\cite{qf-data}. This allows us to not have to manually label experimental data for training purposes.
To prepare the DNNs, we rely on a dataset of $2.7\times10^4$ point fingerprints, sampled over 20 simulated QD devices. 
A number of parameters, such as the device geometry, gate positions, lever arms, and screening lengths, are varied between simulations to reflect the minimum qualitative features across a range of devices. 
For training purposes, each fingerprint $\mathcal{F}_{x_o}$ is tagged with a label identifying the state of the device at point $x_o$.
The labels are generated as part of the simulation.
Before training, the labels are converted to one-hot vectors (i.e., vectors of length equal to the number of classes and a single nonzero element indicating the true class) and treated as the probabilities $\bm{p}(x_o)$ that $x_o$ is in any of the five possible states.

To test the performance of the RBC, we establish an off-line dataset of 311 labeled fingerprints using two measurement scans qualitatively comparable to the one presented in Fig.~\ref{fig:rbc_overview}. 
The points within the test dataset are evenly distributed among the five possible states, with 64 points belonging to the ND class, 58 to the ${\rm SD}_L$ class, 61 to the ${\rm SD}_C$ class, 64 to the ${\rm SD}_R$ class, and 64 to the ${\rm DD}$ class.

Using the fingerprinting configuration as reported in Ref.~\cite{Zwolak20-RBC}, that is, six evenly spaced rays of length 60 pixels (30~\si{\milli\volt}) and a weight function $\gamma(x) = 1/x$ we achieve an average accuracy of 87.1(2.0)~\% ($N=20$ models)~\footnote{We use a notation value(uncertainty) to express uncertainties, for example $1.5(6)\ {\rm cm}$ would be interpreted as $(1.5\pm0.6)~{\rm cm}$. All uncertainties herein reflect the uncorrelated combination of single-standard deviation statistical and systematic uncertainties.}. 
This is better than the 81.9~\% performance for CNN-based classification of experimental data reported in Ref.~\cite{Zwolak20-AQD}. 

\begin{figure*}
\includegraphics[width=1.0\linewidth]{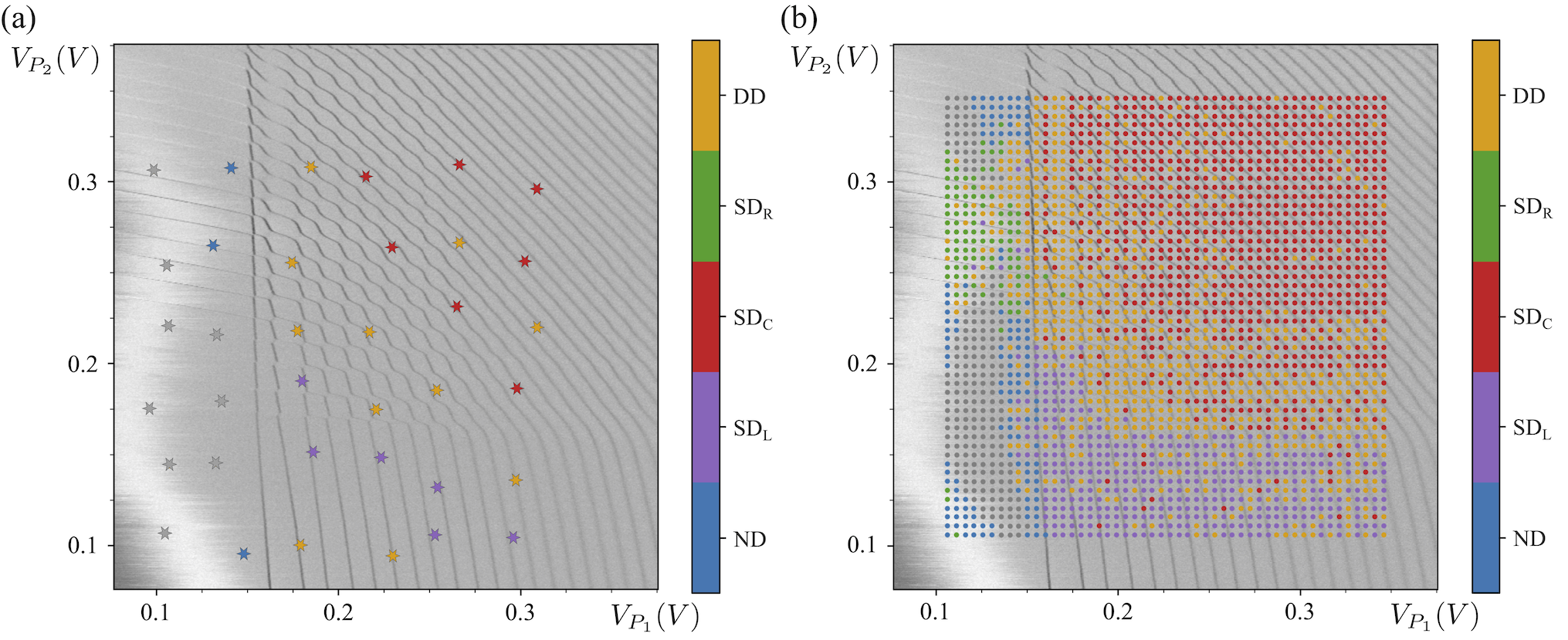}
\caption{Scatter plot showing the performance of the ray-based classifier on live data (a) and off-line (b) using six rays of length 22~\si{\milli\volt} (44 pixels) overlaying a sample measured raw scan. 
Colors on both plots correspond to the state predicted by the RBC. 
The gray points indicate $M$-projections that are determined as poorly charge sensed and thus are inappropriate for classification.
}
\label{fig:rbc_performance}
\end{figure*}

The number of rays, their length, and the choice of the weight function are all considered free parameters of the RBC framework.
To optimize the machine learning process, we start by testing the effect of the weight function on the performance of the classifier.
We use the four most promising combinations of the number of rays and the ray length reported in Ref.~\cite{Zwolak20-RBC}; that is, five and six rays of length 50 pixels (25~\si{\milli\volt}) and of length 60 pixels (30~\si{\milli\volt}).
In our analysis, we consider a collection of three decreasing weight functions with varying decay rates: $\gamma(x)=1/x$, $\gamma(x)=\exp(-x)$, and $\gamma(x)=1-\hat{x}$, where $\hat{x}=(x-\min{\rm {\bf x}})/(\min{\rm {\bf x}}-\max{\rm {\bf x}})$ denotes the min-max normalization. 
In addition, we consider two nondecreasing functions: the min-max normalization $\gamma(x)=\hat{x}$ and the raw distance $\gamma(x)=x$. 
The inset at the top of Fig.~\ref{fig:perf_comparison}(a) shows the performance of the RBC on simulated data.
For $\gamma(x)=\exp(-x)$, the performance is significantly worse than for the other considered functions, averaging at $49(3)~\%$ for six rays and $57(3)~\%$ for 12 rays (for clarity not included in the figure). 
The performance is greatly improved when the argument is min-max normalized, resulting in $95.1(4)~\%$ accuracy for six rays and $96.4(4)~\%$ accuracy for 12 rays.
This suggests that for the non-normalized data the decay rate is too high, making the features indistinguishable for the DNN. 
For completeness, we also consider the min-max normalized version of the function $\gamma(x)=1/x$, finding no difference in performance when compared with the original function [$95.4(4)~\%$ vs $96.7(4)~\%$ for six rays and $94.9(4)~\%$ vs $96.1(4)~\%$ for 12 rays].

Finding no difference in performance when using simulated data, we test all functions using the test set of off-line experimental data.
Figure~\ref{fig:perf_comparison}(a) shows the RBC performance.
While in the absence of noise, all functions considered perform comparably, we find that in the presence of noise, normalization of data with $\gamma(x)=1/x$ consistently leads to significantly better classification accuracy than normalization with the other functions.
For experimental data the performance of the classifier decreases significantly as the weight function rate of change increases. 
For the functions tested, $\gamma(x) =1/x$ has the best balance of sensitivity and robustness against the variability in peak shape and position (see Fig.~\ref{fig:rbc_overview}(b)). 
Additional exploration of different weight functions and peak-finding methods may further improve the performance.

With the measurement efficiency in mind, we also test the effect of the number of rays and their length on the performance. 
We use $M$-projections with $M=5,6,7,9$, and $12$ rays and with lengths ranging between 20 pixels (10~\si{\milli\volt}) and 80 pixels (40~\si{\milli\volt}), sampled every four pixels (2~\si{\milli\volt}).
Since the ray length directly affects the fingerprints (i.e., shorter rays will naturally miss a transition line that would be detected with a longer ray), the rays in the simulated dataset used to train DNNs are adjusted appropriately to ensure compatibility.
As Fig.~\ref{fig:perf_comparison}(b) shows, we find that including more rays does not necessarily lead to greater or more reliable accuracy. 
In addition, for each number of rays considered, there seems to be an optimal length beyond which the performance either stays unchanged or slightly drops until it reaches equilibrium.

To test the RBC {\it in situ}, we develop a measurement routine that enables live acquisition of ray data. 
After selection of a point $x_o$, voltages on gates $P_1$ and $P_2$ are changed in tandem to achieve straight voltage rays emanating from $x_o$. 
This is, in effect, virtual gating of the $(V_{P_1}, V_{P_2})$ voltage space~\cite{Nowack1269, Mills19-CAT}. 
The performance of the classifier for live measurement of 36 points is shown in Fig.~\ref{fig:rbc_performance}(a), with the orientations of the stars indicating the measurement directions.
A quality check on the set of rays is performed before the RBC to prevent classification of poorly charge sensed data [see the changing background signal on the left side of Fig.~\ref{fig:rbc_performance}(a)]. 
The check involves benchmarking of the distribution of voltages measured for a given set of rays---ranging from 120 voltage values for a set of five rays of length 12~\si{\milli\volt} to 720 voltage values for 12 rays of length 30~\si{\milli\volt}---against a threshold established off-line before the experiment based on previously measured rays with clearly discernible charge transitions.
Of the 36 measured points, nine are excluded from classification on the basis of the threshold test. 
The remaining are colored in Fig.~\ref{fig:rbc_performance}(a) according to the class returned by the RBC.
While in the online testing we used $M$-projections with $M=6$ rays of length 22~\si{\milli\volt}, the measurement captured $M=12$ rays of length 40~\si{\milli\volt}.
Additional off-line testing using longer rays does not change the classification results and neither does inclusion of the full 12-rays projections.
This suggests that the protocol used to determine the noise level and signal prominence from real data might require further improvements.
Recalibration of the sensor after each set of rays could also increase the signal-to-noise ratio and lead to more prominent transitions.

\begin{figure*}
    \centering
    \includegraphics[width=1.0\linewidth]{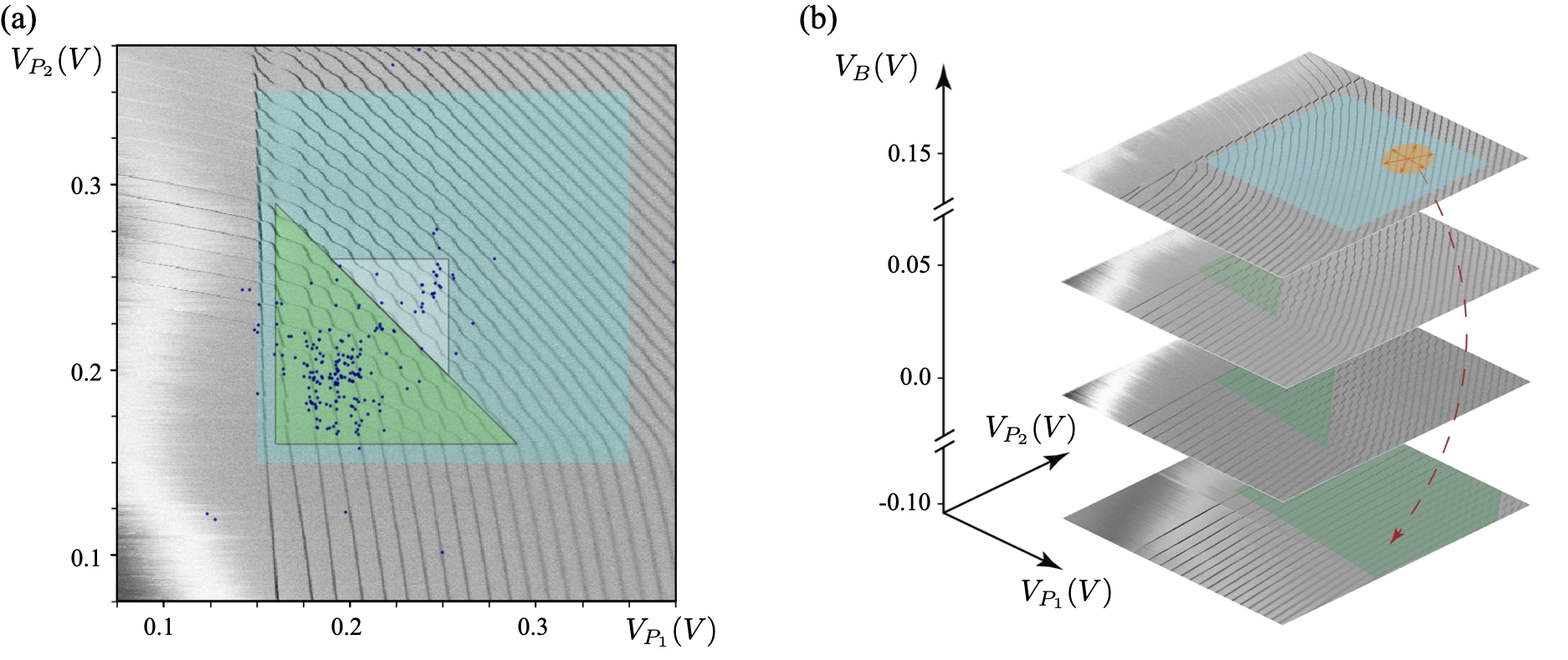}
    \caption{(a) Scatter plot of the final state obtained in off-line tuning using the RBC framework. 
    The tuner is set to tune to the double-quantum-dot state. 
    We obtain a tuning success rate of $78.7\,\%$, calculated as the fraction of points that end up inside the green region, with an  additional $10.2~\%$ of the points landing in an area that moderately resembles double-quantum-dot features, highlighted in white. 
    (b) The three-dimensional space formed by 2D scans of the charge sensor response in the two plunger-gates space and the middle barrier gate.
    The arrow schematically shows the flow of the autotuner in the optimization loop from barrier voltages with no double quantum dot to lower voltages with a double-quantum-dot region.   
    In both plots, the cyan square highlights the area where the initial points are uniformly sampled, while the green polygons mark the target double-quantum-dot region.}
    \label{fig:tuning}
\end{figure*} 

Finally, to assess the performance for a larger set of points, we run the RBC off-line for a set of 2,500 points presampled from a large scan.
The performance is shown in Fig.~\ref{fig:rbc_performance}(b). 
We see that the classifier correctly captures the broad regions in the voltage space that correspond to single QDs---central, left and right---as well as double QD. 
The most common failure cases corresponds to $x_o$ coincidentally lying on the transition lines and in the regions where the lock-in measurement is insensitive (transitions of the $P_2$ QD).
 
%%%%%%%%%%%%%%%%%%%%%%%%%%%%%%%%%%%%%%%%%%%%%%%%%%%%%%%%%%%%%
\subsection{Autotuning with RBC}\label{subsec:rbc-tuning}
%%%%%%%%%%%%%%%%%%%%%%%%%%%%%%%%%%%%%%%%%%%%%
The RBC combined with an optimization loop can be used to tune the device from one state to another (e.g., from single-QD state to a double-QD state). 
We perform off-line tuning by initializing the device at a given point in the space of plunger voltages $x_o=(V_{P_1}$,$V_{P_2})$ and then optimizing a fitness function over a premeasured scan to mimic an actual tuning run. 
The fitness function quantifies how close the probability vector returned by the RBC is to the desired target state. 
We use the fitness function proposed in Ref.~\cite{Kalantre17-MLD,Zwolak20-AQD}:

\begin{equation}\label{eq:fit-func}
    \delta(\bm{p}_{\rm target},\bm{p}(x_o)) = ||\bm{p}_{\rm target} - \bm{p}(x_o)||_2 + \epsilon(x_o) 
\end{equation}
where $||\cdot||$ is the Euclidean norm, $\bm{p}_{\rm target}$ is the probability vector for the target state, $\bm{p}(x_o)$ is the probability vector returned by the RBC at $x_o$, and $\epsilon(x_o)$ is a penalty function for tuning to larger plunger voltages.
We use $\epsilon(x_o) \propto \{\tanh[(V_{P_1} - V_{P_1}^0)/V_0] + \tanh[(V_{P_2} - V_{P_2}^0)/V_0]\}$, where $V_{P_1}^0$ and $V_{P_2}^0$ are previously determined pinch-off values and $V_0$ is a voltage scale normalizing the argument of the $\tanh$ function. 
We use $V_0 = 20\,\si{\milli\volt}$, approximately equal to the charging energy of the QDs. 
The penalty function acts as a regularization function for the bare Euclidean distance between the current and target state probability vectors. 
In particular, it adds a smooth gradient to the background as well as helps the optimizer escape from local minima.

We use the Nelder-Mead optimizer~\cite{Nelder65-NMA} implemented in SciPy~\cite{scipy}. 
The optimizer maintains a set of objective function values at a simplex of $n+1$ points in $n$-dimensional space; in our case it amounts to evaluation on vertices of a triangle in 2D gate space. 
The orientation of the initial simplex is chosen dynamically on the basis of the initial state returned by the RBC and is obtained by changing the voltages on each of the plungers by $40\,\si{\milli\volt}$.
The optimizer works by moving the simplex toward a minimum of the objective function on the basis of the function values at the simplex vertices. 
Since we lack analytic information about the derivative of the fitness function (Eq.~\ref{eq:fit-func}), the Nelder-Mead optimizer is well suited for our purpose as it relies only on function evaluations.

We perform an off-line tuning on a sample premeasured large 2D scan to test the viability of the RBC framework in tuning the device state. 
The final state to be tuned to is set to the double-QD state. 
The initial points are uniformly sampled in a square grid over a range of 200~\si{\milli\volt}, which encompasses approximately 18 electron transitions [highlighted in cyan in Fig.~\ref{fig:tuning}(a)].
During the tuning loop, the rays are sampled at each point by linear interpolation within the 2D scan on a grid.
Fig.~\ref{fig:tuning} shows a scatter plot of the final state at the end of the tuning loop.
To quantify the performance, we define a triangular region [highlighted in green in Fig.~\ref{fig:tuning}(a)] as the success region for tuning to a double-QD state. 
We report a tuning success rate of $78.7\,\%$ for a set of 225 uniformly sampled initial points, with an additional $10.2~\%$ of the points landing in an area that moderately resembles double-QD features.
For comparison, the success rate for tuning the 2D scans reported in Ref.~\cite{Zwolak20-AQD} is $75(32)~\%$ when the tuning is started from a region enclosing at most nine transition lines.  

Finally, we perform off-line tuning in a three-dimensional (3D) space formed by a series of scans in the plunger gates space taken at different values of the middle barrier gate. 
As can be seen in Fig.~\ref{fig:tuning}(b), by varying the middle barrier from $-100$ to $150\,\si{\milli\volt}$, the device can be tuned from having predominantly double-QD features to having predominately single-QD features.
The green overlays on the scans in Fig.~\ref{fig:tuning}(b) highlight the double-QD regions.
For reference, the scan used in Fig.~\ref{fig:tuning}(a) is taken with the middle barrier set to $50\,\si{\milli\volt}$. 
The rays at a given point $(V_{P_1},V_{P_2},V_B)$ are sampled as before in the plunger space, but the fitness function now includes $V_B$ in its argument. 
We initialize 100 tuning runs within the top scan, as highlighted in cyan in Fig.~\ref{fig:tuning}(b), and tune to the double-QD state, finding an overall success rate of $67\,\%$ for tuning in three dimensions. 

The failure modes for the tuning process in both two dimensions and three dimensions include landing at transition lines where the fingerprint does not correspond to either a single-QD state or double-QD state as well as converging to local minima of the fitness function. 
Although the addition of regularization $\epsilon(x_o)$ mitigates the latter to some extent, further work on optimization algorithms is necessary to increase the tuning success rate.
Incorporation of a CNN-based classifier to verify the state of the final state and, if necessary reinitiate the autotuner, would likely help alleviate the former issue. 
In comparison with the tuning results reported with CNNs~\cite{Zwolak20-AQD,Kalantre17-MLD}, the RBC framework requires a comparable number of iterations to achieve the same end goal, leading to a significant reduction in data acquisition (approximately $60\,\%$) with use of rays instead of 2D scans.

%%%%%%%%%%%%%%%%%%%%%%%%%%%%%%%%%%%%%%%%%%%%%%%%%%%%%%%%%%%%%
\section{Summary and Conclusions}\label{sec:conclusion}
%%%%%%%%%%%%%%%%%%%%%%%%%%%%%%%%%%%%%%%%%%%%%%%%%%%%%%%%%%%%%
To summarize, in this paper, we discuss an experimental implementation of the ray-based classification framework using double-quantum-dot devices.
We propose a measurement scheme relying on one-dimensional projections in the plunger gates space as means to ``fingerprint'' the device states.
With measurement efficiency in mind, we consider various combinations of the number of rays and the length of rays as well as multiple weight functions to determine an optimal balance between measurement load and classification accuracy.
We show that for the device used, the performance accuracy remains at about $87~\%$ regardless of whether six, seven, or nine rays are used.
This translates to an up to approximately $70~\%$ reduction in the number of measured points needed for classification compared with the CNN-based approach.
Increasing the number of rays to 12 results in an accuracy of about $90~\%$, while reducing the number of points measured by $40~\%$ (see Table~\ref{tab:cost_perf} for comparison of all ray numbers tested).

\begin{table}[t]
\renewcommand{\arraystretch}{1.1}
\renewcommand{\tabcolsep}{6pt}
\caption{Summary of the performance for five $M$-projections for rays of length length $12$ and $22~\si{\milli\volt}$. The accuracy is averaged over $N=20$ models and the data reduction $\Delta$ indicates the expected reduction in the number of measured points needed for classification compared with the CNN-based approach.} 
\centering
\begin{tabular}{p{0.4cm}p{2cm}p{0.9cm}p{0.02cm}p{2cm}p{0.9cm}}
\hline \hline
& \multicolumn{2}{c}{12~\,\si{\milli\volt}} & &\multicolumn{2}{c}{22~\,\si{\milli\volt}} \\
\cline{2-3}\cline{5-6}
$M$ & Accuracy (\%) & $\Delta (\%)$ &&  Accuracy (\%) & $\Delta$ (\%)\\
\hline
 $5$ & 79.4(3.3) & 87 && 84.4(2.1) & 76 \\
 $6$ & 82.7(1.2) & 84 && 87.4(1.6) & 71 \\
 $7$ & 83.2(3.1) & 81 && 86.4(3.4) & 66 \\
 $9$ & 83.1(1.7) & 76 && 87.1(1.4) & 56 \\
$12$ & 81.1(1.3) & 68 && 89.8(0.9) & 41 \\
\hline \hline
\label{tab:cost_perf}
\end{tabular}
\end{table}

We also show how the RBC framework can be implemented to tune the QD device in 2D and 3D gate space. 
We perform autotuning on a series of premeasured scans in 2D and 3D gate voltage spaces, reliably tuning the device from one state to another.
In this work, we focus on automated tuning of a QD device into a voltage space with coupled double QDs. 
It is also important to note that this tuning scheme does not achieve a specific occupation of each QD, but rather achieves a few-electron double-QD regime. 
Depending on the intended functionality, (single-electron qubit, multielectron qubit, etc), additional methods are required to achieve an exact occupation for each QD. 

With the noisy intermediate-scale quantum technology era on the horizon~\cite{Preskill2018NISQ}, it is important to consider the practical aspect of implementing automated control as part of the device itself, in the ``on-chip'' fashion.
The network architecture necessary for RBC is significantly simpler and smaller than for CNN-based classification, making it more suitable for an implementation on miniaturized hardware with low power consumption in the near future~\cite{Sebastian20-MIC,Czischek21-MNN}.
In particular, the neural network used to train the RBC comprises only four fully connected dense layers with 128, 64, 32, and 5 units, respectively.
The total number of parameters necessary for the RBC is about $1.2\times10^4$  (compared with about $2.2\times10^6$ parameters defining the CNN-based model in Ref.~\cite{Zwolak20-AQD})
An optimization of the network architecture was not considered here. 
However, from our preliminary inspection, we expect that it will be possible to further reduce the number of parameters defining the model by an order magnitude.

With increasing complexity of QD devices in both QD number and gate geometry, the need for automated state identification and tuning will increase. 
With the development of QD-based spin qubits using industrial technologies \cite{veldhorst2017silicon}, a technique that enables efficient and scalable characterization of QDs for qubit applications is necessary. 
The RBC framework introduced in this work is a measurement-cost-effective solution for state classification and tuning. 
Further improvements to RBC include using features beyond the critical feature, nonuniform distribution of rays, or changes to the weight function. 
While the RBC framework is only a piece of the full device tuning puzzle, it is a powerful and efficient tool toward quantum dot automation.

%%%%%%%%%%%%%%%%%%%%%%%%%%%%%%%%%%%%%%%%%%%%%%%%%%%%%%%%%%%%%
\begin{acknowledgements}
We acknowledge Lisa Edge from HRL Laboratories for the growth and distribution of the Si/Si$_x$Ge$_{1-x}$ heterostuctures that are used in the experiment.
This research was sponsored in part by the U.S. Army Research Office through Grant No. W911NF-17-1-0274. 
S.S.K. gratefully acknowledges support from the Joint Quantum Institute--Joint Center for Quantum Information and Computer Science  Lanczos Graduate Fellowship. 
We acknowledge the use of clean room facilities supported by the NSF through the University of Wisconsin-Madison Materials
Research Science and Engineering Center (Grant No. DMR-1720415) and electron beam lithography equipment acquired with support of the NSF Major Research Instrumentation Program (Grant No. DMR-1625348). 
The views and conclusions contained in this paper are those of the authors and should not be interpreted as representing the official policies, either expressed or implied, of the U.S. Army Research Office, or the U.S. Government. 
The U.S. Government is authorized to reproduce and distribute reprints for U.S. Government purposes notwithstanding any copyright noted herein. 
Any mention of commercial products is for information only; it does not imply recommendation or endorsement by the National Institute of Standards and Technology.
\end{acknowledgements}
%%%%%%%%%%%%%%%%%%%%%%%%%%%%%%%%%%%%%%%%%%%%%%%%%%%%%%%%%%%%%

%%%%%%%%%%%%%%%%%%%%%%%%%%%%%%%%%%%%%%%%%%%%%%%%%%%%%%%%%%%%%%

%%%%%%%%%%%%%%%%%%%%%%%%%%%%%%%%%%%%%%%%%%%%%%%%%%%%%%%%%%%%%%
\end{document}